\documentclass[%
 reprint,
superscriptaddress,
showpacs,preprintnumbers,
 amsmath,amssymb,
 aps,
floatfix,
]{revtex4-1}
\usepackage{mathtools}
\usepackage{graphicx}
\usepackage{dcolumn}
\usepackage{bm}
\usepackage{hhline}
\usepackage{color,soul}
\usepackage{float}
\usepackage{mathrsfs}

\begin{document}


\title{A nucleation theory for yielding of nearly defect-free crystals: understanding rate dependent yield points}

\author{Vikranth Sagar Reddy}
\affiliation{%
Tata Institute for Fundamental Research,  Centre for Interdisciplinary Sciences, 36/P Gopanapally, Hyderabad 500107,  India.
}%
\author{Parswa Nath}%
\affiliation{%
Tata Institute for Fundamental Research,  Centre for Interdisciplinary Sciences, 36/P Gopanapally, Hyderabad 500107,  India.
}%

\author{J\"urgen Horbach}
\affiliation{Institut f\"ur Theoretische Physik II: Weiche Materie, Heinrich 
Heine-Universit\"at D\"usseldorf, Universit\"atsstra{\ss}e 1, 40225 D\"usseldorf, 
Germany}
\author{Peter Sollich}
\affiliation{Institute for Theoretical Physics, Georg-August-University G\"ottingen, 37077 G\"ottingen, Germany 
}%
\affiliation{%
Department of Mathematics, King's College London, Strand, London WC2R 2LS, UK.}
\author{Surajit Sengupta}
\affiliation{%
 Tata Institute for Fundamental Research,  Centre for Interdisciplinary Sciences, 36/P Gopanapally, Hyderabad 500107,  India.
}%

\date{\today}
\begin{abstract}
Experiments and simulations show that when an initially defect free rigid crystal is subjected to deformation at a constant rate, irreversible plastic flow commences at the so-called {\em yield point}. The yield point is a weak function of the deformation rate, which is usually expressed as a power law with an extremely small non-universal exponent. We re-analyze a representative set of published data on nanometer sized, mostly defect free, Cu, Ni and Au crystals in the light of a recently proposed theory of yielding based on nucleation of stable stress-free regions inside the metastable rigid solid. The single relation derived here, which is {\em not} a power law, explains data covering {\em fifteen} orders of magnitude in time scales.     
\end{abstract}
\maketitle
The phenomenon of yielding is possibly the most conspicuous, and therefore well studied, aspect of the mechanical response of materials upon external loading~\cite{rob, argon, rei, sethna2017}. In a typical experiment a sample of material is deformed at a constant rate and the point at which the stress suddenly drops, in a marked departure from reversible elastic behavior, is identified with the yield point~\cite{rei}. No theory, starting from rigorous thermodynamic principles (e.g.~shape independence of equilibrium free energies)~\cite{ruelle}, has been derived that predicts the yield point of a real solid. Detailed phenomenological models, on the other hand, have been extremely useful for this purpose~\cite{rob, argon}. All existing work on irreversible mechanical processes in crystalline solids ultimately connects to the static and dynamic properties of lattice defects viz.\  dislocations~\cite{hirth}. These are known to 
interact with each other as well as with impurities, grain boundaries, pinning sites etc.~\cite{rob, argon}. However, a quantitative understanding of the collective behavior of large accumulations of dislocations, readily observed in large scale atomistic computer simulations~\cite{yip} and expected to play a key r\^ole in yielding, remains elusive. Various models, such as the thermodynamic dislocation theory~\cite{TDT}, based on de-pinning, ~\cite{depin} or un-jamming~\cite{jamming} of dislocations have been proposed. Yielding is also viewed as a dynamical critical phenomenon~\cite{sethna2017,CL}.  Indeed, scale-free avalanches and intermittent phenomenon have been observed to accompany yielding~\cite{avalanche}. A non-zero initial dislocation concentration is indispensible for most known models of yielding.  
\begin{figure}[ht]
 \begin{center}
\includegraphics[scale=0.18]{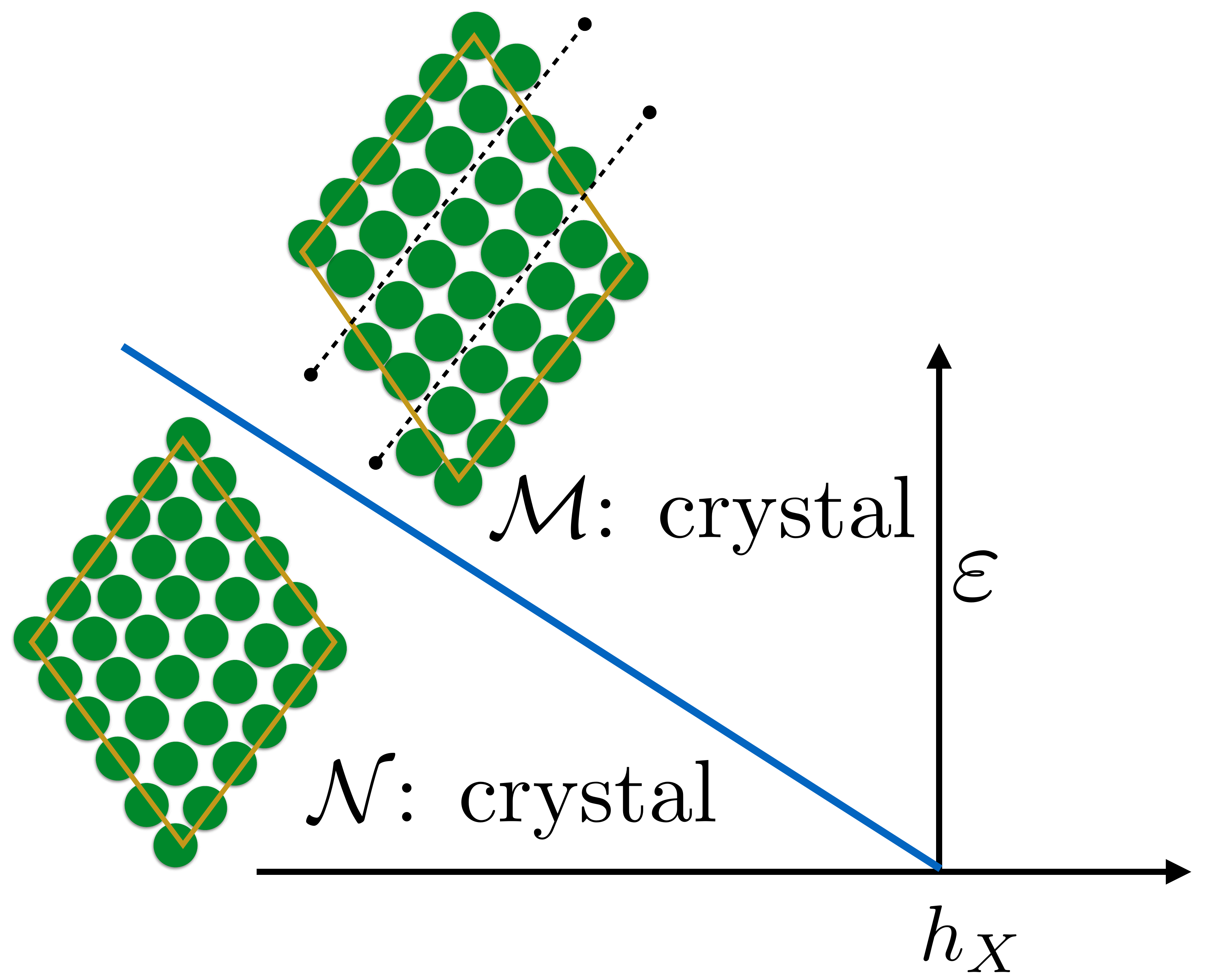}
\caption{Schematic phase diagram and $\mathscr{N}$, $\mathscr{M}$ phases~\cite{pnas} corresponding to the $d=2$ square lattice, in the $h_X - \varepsilon$ plane (see text). The blue line is the first order phase boundary. The dotted lines show slip planes in the $\mathscr{M}$ phase.}
\label{schema}
\end{center}
\end{figure}

\begin{table*}[]
\begin{tabular}{|l|c|c|c|c|c|c|c|c|}
\hline
Ref.     & Material & Method & Details & K (GPa) & $\tau_0$ (fs) &$\gamma_s$ (mJ/m$^2$) &$\alpha^{\frac{1}{4}}$ \\ \hhline{|========|}
\cite{jenn} &  Cu  & Expt. : compressive& $\approx$500nm diameter single crystal nano-pillars&130&39.23&1.826&0.0096 \\ \hline
\cite{cheng} & Cu  & Expt.: tensile &poly-crystal with $\approx$30nm grains&130&37.59&2.801&0.0120 \\ \hline
\cite{peng} & Ni  & Expt.: tensile&$\approx$200-300nm diameter nano-wires&200&57.55&4.207&0.0131 \\ \hline
\cite{cholla} & Ni  & Expt.: compressive&poly-crystal with $\approx$100-1000nm grains&200&86.82&3.138&0.0106  \\ \hline
\cite{cholla} & Ni  & Expt.: compressive&poly-crystal with $\approx$40nm grains&200&34.12&6.040&0.0173  \\ \hline
\cite{gu} & Ni & Expt.: compressive&poly-crystal with $\approx$20nm grains&200&44.42&7.662&0.0206 \\ \hline
\cite{hump} & Ni  & Expt.: tensile& poly-crystal with $\approx$30nm grains&200&35.43&1.341&0.0056 \\ \hline
\cite{park} & Au  & MD: tensile&5.176$\times$5.176nm nano-wire&188&24.05 &9.536&0.0251 \\ \hline
\cite{shi} & Au  & MD: tensile (0$^\circ$C)&5.65nm diameter nano-wire&188&341.34&18.367&0.0420\\ \hline
\cite{spju} & Au   & MD: tensile (-269$^\circ$C)&1nm diameter nano-wire&188&11.02&4.658&0.0432 \\ \hline
\cite{wen} & Ni  & MD: tensile&2.53nm diameter  nano-wire&200&63.87&35.141&0.0648 \\ \hline
\cite{hawu} & Cu  & MD: tensile &2$\times$2nm nano-wire&130&312.84&39.718&0.0880 \\ \hline
\cite{key} & Cu (1) & MD: tensile&9.18$\times$9.18nm nano-wire&130&36.59&3.051&0.0128 \\ \hline
\cite{key} & Cu (2)& MD: tensile&9.18$\times$9.18nm nano-wire&130&32.97&3.100&0.0129 \\ \hline

\end{tabular}
\caption{
The results of nonlinear least squares fits of Eq.~(\ref{lambert}) to $\tau_0$ and $\alpha$. Details of the method, i.e.~compressive or tensile loading and the materials considered, are also given along with the references (column 1). The values of $K$ are taken from~\cite{props} (Cu, Ni), and~\cite{park} (Au). Except where mentioned otherwise, data correspond to room temperature (27$^\circ$C). The interfacial energy $\gamma_s$ is computed from $\alpha$ and $K$. 
The last two rows give parameters obtained by fitting the same data using two different methods. Typical errors from MD are $\sim \pm 20\%$ for $\tau_0$ and $\lesssim \pm 10\%$ for $\gamma_s$; error estimates for experimental data are higher due to more scatter and a smaller range of ${\dot \varepsilon}$ (see~\cite{supplm}). }
\label{table}
\end{table*}

 One of the most intriguing features of yield phenomena in solids is the history dependence of the yield point. It is known that the yield point stress, $\sigma_Y$, is a weak function of the rate, $\dot \varepsilon$, by which the solid is deformed {\em prior} to the commencement of irreversibility~\cite{rob, argon, johnson-cook, borodin, meyers}. This relation is expressed as a phenomenological power law $\sigma_Y \sim (\dot \varepsilon)^m$. The rather non-universal, strain rate sensitivity (SRS) exponent $m$ varies over a wide range $0.006 - 0.06$ depending on many factors such as the nature of the solid, the size of the specimen as well as $\dot \varepsilon$ itself. The weak dependence on $\dot \varepsilon$ and the small value of $m$ may be the reason why this aspect of yielding has not received the attention that it deserves. Usually very low (or very high) values of fitted exponents point to a gap in our understanding of the underlying physics, hinting that the actual relation may not, in fact, be a power law. 
 
In this Letter we show that, at least for one special class of
materials\textcolor{black}{, viz.~}nearly defect-free crystals, this
apprehension is true. We derive and test a relation between the yield
strain $\varepsilon^*$ defined simply as $\sigma_Y/K$, where
$K$ is the appropriate elastic modulus, and $\dot \varepsilon$. This
relation is {\em not} a power law and involves instead an essential
singularity. It explains experimental and computer simulation data over
a very wide range of temporal scales.

\begin{figure*}[ht]
 \begin{center}
\includegraphics[scale=0.27]{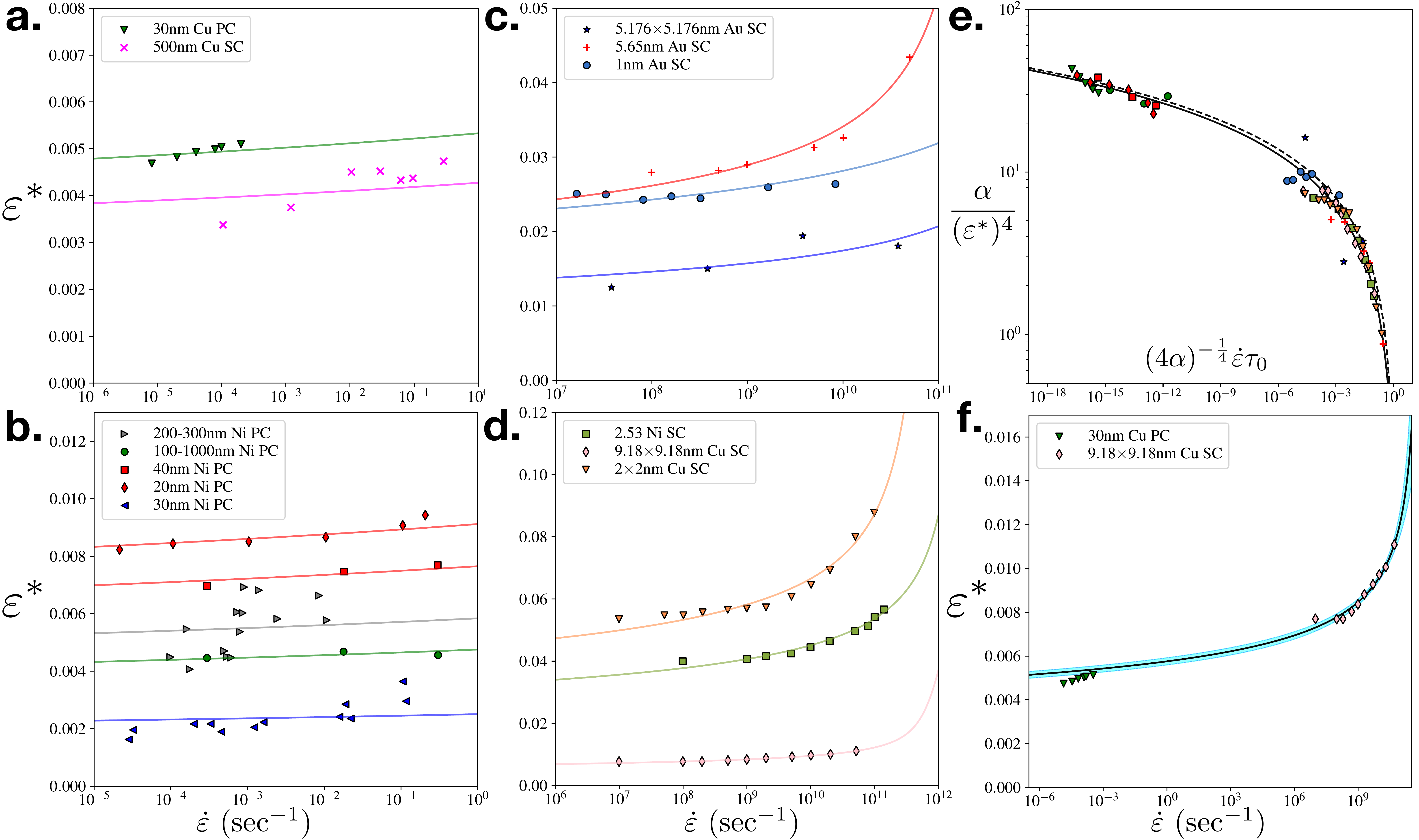}
\caption{{\bf a},{\bf  b.} Non-linear least square fits (solid lines) of Eq.~(\ref{lambert}) to experimental results (symbols) on Cu and Ni samples, both single crystal (SC) and ultra fine grained polycrystalline (PC); see text and Table~\ref{table} for details. {\bf c},{\bf d.} Same as {\bf a},{\bf b} but for MD simulations of Au, Cu and Ni. {\bf e.} Plot of the data in {\bf a}--{\bf d} using scaled variables (with the same symbols as in {\bf a}--{\bf d.}) showing collapse onto the single master curve~(\ref{lambert}) (solid black line). Dashed: Approximate asymptotic form~(\ref{lambert_asympt}). Three experimental data sets with excessive ($ > 50\%$) scatter have been omitted. {\bf f.}\ A log-linear plot of Eq.~(\ref{lambert}) shows that the parameters $\gamma_s$ \& $\tau_0$ obtained from MD of Cu~\cite{key}, are able to predict the results of experiments~\cite{cheng} performed $10^{-12}$ times more slowly. The light blue band on the solid line shows the uncertainty of our prediction at $95\%$ confidence (from fit (2), see Table~\ref{table}).}
\label{univ}
\end{center}
\end{figure*}

Our primary idea emerges from the following, somewhat surprising, fact. Even though $\sigma_Y$ appears to decrease only slightly when $\dot \varepsilon$ decreases by a large amount, {\em crystalline solids are guaranteed to yield at infinitesimal stresses when deformed at vanishing rates.} This conclusion follows from very general considerations starting with the exact result that the free energy of any material, made up of entities interacting with short ranged forces, cannot depend on the shape of the boundary~\cite{ruelle, penrose}. This seemingly pedantic issue can be converted into a quantitative calculation by showing that even at vanishingly small strains an initially dislocation free ideal crystal may nucleate regions, or bubbles, inside which atoms rearrange to eliminate stress~\cite{sausset}. These stress free bubbles are embedded inside the deformed and stressed solid and bounded by an interface containing an array of defects. Since the volume energy contribution of the bubble is always negative, the surface energy cost from the defect array can be compensated as the size of the bubble increases. A rigid solid is therefore always metastable at infinitesimal deformation, although the relaxation time for bubble nucleation diverges as $\varepsilon \to 0$. 
 
Recently, by introducing a fictitious field, $h_X$, that penalizes atomic rearrangements, some of us showed  that further insight may be obtained into this problem~\cite{pnas} (see also Supplementary Material~\cite{supplm}). One can then analyze the problem in the 
full $h_X - \varepsilon$ plane in the thermodynamic limit and let 
$h_X \to 0$ at the end. This work concludes that there exists an 
equilibrium first order transition~\cite{pnas} between two kinds 
of ideal crystal, say $\mathscr{N}$ and $\mathscr{M}$ in the $h_X - \varepsilon$ 
plane with a phase boundary that extrapolates to $\varepsilon = 0$ 
at $h_X = 0$ (Fig.~\ref{schema}). The two crystals have identical symmetry 
but differ in the way they respond to deformation. The $\mathscr{N}$ 
crystal produces an internal restoring stress upon deformation, while 
the $\mathscr{M}$ crystal undergoes spontaneous rearrangements 
(e.g.\ by slipping) to accommodate the deformation at zero stress. The 
latent heat released during the $\mathscr{N}\to\mathscr{M}$ transition is just 
the elastic energy stored in the rigid, $\mathscr{N}$, phase. For any 
$h_X = 0$ and $\varepsilon \neq 0$, the $\mathscr{N}$ phase is metastable and 
decays to the stable $\mathscr{M}$ phase by bubble nucleation. 
Then, for a given $\varepsilon$, the free energy $\mathscr{F}$ 
of a spherical bubble of $\mathscr{M}$ phase, with radius $R$ and in $d$ 
dimensions, is given within a linear approximation by
$\mathscr{F} = -\frac{1}{2}K\varepsilon^2 V_1 R^d + \gamma_s S_1R^{d-1}$.
Here, $K$ is the elastic modulus and $V_1 = {\pi^{\frac{d}{2}}}/{\Gamma(\frac{d}{2}+1)}$ 
and $S_1 = d V_1$ are the volume and surface area of a $d$ dimensional unit sphere, 
respectively. The equilibrium surface energy, $\gamma_s$ is rigorously defined only at 
coexistence at the first order boundary~\cite{tfot}. Extremizing $\mathscr{F}$ gives 
the free energy barrier $\Delta \mathscr{F}$, at a critical value of the radius $R = R_c$, as
$$\Delta \mathscr{F} = 
[2(d-1)]^{d-1} V_1 \gamma_s^d K^{-(d-1)}
\varepsilon^{-2(d-1)}.$$
Up to now, we have considered the limit $\dot{\varepsilon}\to 0$. At a finite strain
rate, the strain depends on time $t$, i.e.~$\varepsilon=\dot{\varepsilon} t$, and 
thus the free energy barrier becomes time-dependent, 
$\Delta \mathscr{F} = \Delta \mathscr{F}(\dot{\varepsilon} t)$.
The time $\tau_{FP}$, at which the system yields, can be understood as a (first passage)
barrier crossing time
and is associated with a critical strain $\varepsilon^*=\dot{\varepsilon} \tau_{FP}$.
Thus, it is given by~\cite{becker} 
$\tau_{FP}\equiv \varepsilon^*/\dot{\varepsilon} = \tau_0 \exp(\beta \Delta \mathscr{F}(\varepsilon^*))$,
where $\beta$ is the inverse temperature and $\tau_0$, the inverse ``attempt frequency'', is 
a microscopic timescale that is independent of the barrier. The equation for $\tau_{FP}$ is a self-consistency equation for $\varepsilon^*$~\cite{sc-cnt}. Here, $\beta \Delta \mathscr{F} = \alpha (\varepsilon^*)^{-2(d-1)}$ with
$\alpha = \beta [2(d-1)]^{d-1} V_1 \gamma_s^d K^{-(d-1)}$ and thus there is an essential singularity
in the dependence of  $\tau_{FP}$ on $\varepsilon^*$.

In Ref.~\cite{pnas} a similar equation, derived for $d=2$, was solved numerically. Here we show that it is possible to solve the relation analytically and obtain a closed form expression for $\varepsilon^*$ 
as a function of $\dot \varepsilon$. Defining
$x=2(d-1) \alpha (\varepsilon^*)^{-2(d-1)} = 2(d-1)\beta \Delta\mathscr{F}$ we can write 
$2(d-1)\alpha(\dot \varepsilon \tau_{FP})^{-2(d-1)} = x\exp(x)$. This can now be inverted using the 
Lambert W function~\cite{why-w}: if $x e^x = f(x) = y$ then $x = f^{-1}(y) = W(y)$. We thus have 
finally
\begin{equation}
\varepsilon^* =
\left[\frac{W\left(2(d-1)\alpha(\dot \varepsilon \tau_0)^{-2(d-1)}\right)}{2(d-1) \alpha }\right]^{-1/[2(d-1)]}
\label{lambert}
\end{equation}
This is the desired closed form expression for the yield strain, a smooth and monotonically increasing function of $\dot \varepsilon$. Note that there are only two free parameters in Eq.~(\ref{lambert}), $\tau_0$, a characteristic time and $\gamma_s$ (or alternately $\alpha$), a characteristic energy. Also, wherever 
Eq.~(\ref{lambert}) is valid, $\tau_0$ is a constant. In Ref.~\cite{pnas}, this was tested for a $d=2$ perfect, initially defect-free Lennard-Jones solid~\cite{CL}. First  $\gamma_s$ was computed at phase coexistence using advanced Monte Carlo sampling methods for nonzero $h_X$ and $\varepsilon$, 
with the final $h_X \to 0$ limit taken by numerical extrapolation. Eq.~(\ref{lambert}) was then solved for $\tau_0$ using $\epsilon^*$ obtained from molecular dynamics (MD) at various $\dot \varepsilon$. The results showed that this theory was excellent for $\beta \Delta \mathscr{F}\gtrsim 1$.

By construction, of course, Eq.~(\ref{lambert}) is thermodynamically consistent. Can it also explain real experimental data in $d = 3$? We check this here by re-analyzing a representative, though not exhaustive, set of available data ~\cite{jenn,peng,park,shi, spju, cholla, wen, key, gu, cheng, wei, hump, hawu}. The systems chosen consist of nearly defect-free ductile single crystals of high purity elemental metals like Cu~\cite{jenn, cheng} and Ni~\cite{peng,cholla,gu,hump}. We include not only nano-pillars~\cite{jenn} and nano-wires~\cite{peng} but also poly-crystalline samples~\cite{cheng,cholla,gu,hump} with ultra fine grains. Lastly, we also re-analyze the results of large scale MD simulations~\cite{park,shi,spju,wen,key,hawu} of Au, Cu and Ni single crystals within our framework. Ideally, $\tau_0$ and $\gamma_s$ should depend only on the material, but in practice factors like the experimental protocol e.g.~tensile or indentation, sample size, proximity to free surfaces, pre-stress, frozen-in defects etc.\ may be important, causing variation in our fit parameters.

For all these systems (see~\cite{supplm} for a summary), reasonably accurate yield point stresses $\sigma_Y$ (under either compressive nano-indentation or tensile loads) have been reported as a function of \textcolor{red}{$\dot \varepsilon$}. The original authors of Refs.~\cite{jenn,peng,park,shi, spju, cholla, wen, key, gu, cheng, wei, hump, hawu} had fitted the data to power laws and extracted widely different values of the SRS  exponent $m$. We first digitize the available data and divide the quoted yield stress, $\sigma_Y$, by the appropriate elastic modulus of the bulk solid~\cite{props} to obtain $\varepsilon^*$; this is a dimensionless quantity that is easier to compare among the different data sets. Fits to Eq.~(\ref{lambert}) then give values of $\alpha$ and $\tau_0$. The fits are shown in Fig.~\ref{univ} {\bf a}--{\bf d}; Table~\ref{table} lists the corresponding parameter values. Eq.~(\ref{lambert}) describes the data rather well, with some of the experimental data showing, understandably, more scatter than the MD ones. The fitted values are also remarkably consistent across the different sets of data, apart from a few exceptions coming mainly from MD studies of very small crystals, where finite size effects may be strong. Most values for $\tau_0$ ($\sim 10-100~{\rm fs}$) are in line with typical dislocation nucleation times reported earlier~\cite{tao} as well as our estimates based on MD simulations~\cite{pnas}. The interfacial energy, $\gamma_s$ ($\sim 10$ mJ/m$^2$), computed from $\alpha$, is also consistent with that of low energy solid interfaces (e.g. stacking faults)~\cite{rob}. The assumption of homogeneous nucleation of the stress-free $\mathscr{M}$ state, implicit in Eq.~(\ref{lambert}), does not affect the qualitative nature of the solution as long as $\gamma_s > 0$. Systemic variations of $\gamma_s$ are difficult to resolve, however, given the scatter in the original data.

We next look in more detail at the strain rate dependence of the yield strain predicted by Eq.~(\ref{lambert}). For small strain rates, where the argument of $W(y)$ becomes large, one can use the asymptotic behaviour $W(y)=(\ln y)-\ln(\ln y)+\ldots$ The yield strain thus has a logarithmic rather than power law dependence on strain rate, which rationalizes the small effective SRS exponents found previously. Keeping only the leading term $W(y)\approx \ln y$ gives, in $d=3$, 
\begin{equation}
    \alpha (\varepsilon^*)^{-4} = W(4\alpha(\dot \varepsilon \tau_0)^{-4})/4 \approx \ln[(4 \alpha)^{1/4} (\dot \varepsilon \tau_0)^{-1}]
\label{lambert_asympt}
\end{equation}
These relations (solid and dashed lines, respectively) are plotted along with experimental and simulation results in Fig.~\ref{univ}{\bf e}, showing collapse of data sets covering, altogether, fifteen orders of magnitude in time scale. Both forms, involving either $W(y)$ or $\ln(y)$, describe the data extremely well in this range. This makes sense as with $\tau_0 \sim$ 30 fs ($3\times 10^{-14}$s) our data lie in the range $({\dot \varepsilon \tau_0})^{-1} \sim 10^3\ldots10^{19}$ and $4\alpha(\dot \varepsilon \tau_0)^{-4} \sim 10^{4} \ldots 10^{68}$
is indeed typically large.
The dimensionless nucleation barrier $\beta \Delta \mathscr{F}$ shown on the y-axis is considerably larger for the experiments than for the MD data, where the barrier heights are at the limit of validity of our theory.

We find it remarkable that Eq.~(\ref{lambert}) is able to explain such a large and varied set of data. It is particularly surprising that even data from ultra fine grained poly-crystalline samples appears to be describable within our picture.  In these systems, there are no dislocations within the grains since the grain boundaries function as dislocation sinks~\cite{rob,argon}. The entire sample therefore behaves as an ensemble of nearly independent crystallites. 

Defect free, nanometer sized crystals are an important group of materials that are fast gaining technological relevance~\cite{caobook}. Our idea, based on a simple, classical, picture of nucleation of stress-free bubbles of $\mathscr{M}$ inside $\mathscr{N}$ describes yielding and SRS in these systems. 
In our picture, dislocations occur only at the interface between the $\mathscr{N}$ and $\mathscr{M}$ phases in order to match the two lattices. Their energy is {\em subsumed} into $\gamma_s$ and they are constrained (by definition) to always move together with the interface, where the local stress gradient is large~\cite{amit-str}. 
Interfaces exist either at equilibrium phase co-existence or when the growth of the $\mathscr{M}$ phase is arrested due to kinetic effects. In real situations ($h_X = 0$) phase co-existence does not occur for nonzero deformation in the thermodynamic limit~\cite{pnas}. Observed dislocations are associated, therefore, with non-equilibrium, or kinetically arrested, configurations. In experiments, dislocations are far easier to see directly, using electron microscopy~\cite{rob}, than the stress interface itself. The notion of a first order transition has, therefore, escaped attention. 

The two parameters of our theory, $\tau_0$ and $\gamma_s$ (or $\alpha$) can be obtained by either fitting yielding data, as is done here, or from precise, finite sized scaled, numerical computations as in Ref.~\cite{pnas}. We emphasize that our nucleation theory respects rigorous thermodynamic constraints~\cite{ruelle} and {\em at the same time} offers a way to successfully predict ``real-world" experimental data at low strain rates from simulations performed at much higher rates. We demonstrate this in Fig.~\ref{univ}{\bf f} by using parameters obtained from the MD results of Ref.~\cite{key} to predict 
accurately the outcome of the experiments of Ref.~\cite{cheng}, which are separated from the simulation data by twelve orders of magnitude in time. This is especially remarkable because an alternative form for $\Delta\mathscr{F}$ derived in Ref.~\cite{sausset} without considering a first order phase transition obtains qualitatively similar $\varepsilon^*({\dot \varepsilon})$ but physically unreasonable yield points~\cite{supplm}.  

Finally, we believe that the work reported here allows us to re-iterate an interesting perspective on the phenomenon of yielding. Despite the success of Eq.~(\ref{lambert}) in explaining such an extensive set of data, it is not a ``theory for everything". Yielding in initially defect free crystals is {\em metastability controlled}, with the large barrier, $\Delta \mathscr{F} > k_B T$, between the distinct $\mathscr{N}$ and $\mathscr{M}$ phases playing a dominant role in determining the dynamical response of the solid to external load. On the other hand, in a solid with a large number of frozen-in defects, the barrier may be much smaller and may even vanish, $\Delta \mathscr{F} \to 0$. There is also a possibility of many co-existing, relatively shallow minima in $\mathscr{F}$ appearing~\cite{supplm}.
If barriers disappear, the nucleation route could be replaced by spinodal decomposition~\cite{CL}, with accompanying critical-like behavior, avalanches etc.\ as is observed~\cite{avalanche}. In this regime, the yielding process would be {\em instability controlled}~\cite{instab}. An actual dynamical critical point is also possible~\cite{sethna2017, depin, jamming}. The situation in amorphous solids may be similar; here yielding is known to be preceded by the percolation of mobile clusters and critical behavior is often noticed~\cite{glasyield}. Our theory cannot predict yielding behavior in this regime, where one commonly obtains either zero or even negative values of the SRS exponent $m$~\cite{negative-m}. Is it possible to develop a theory of yielding with more general validity, capable of describing both the nucleation and critical regimes? We hope that our work generates enthusiasm for pursuing this direction of future research.    

We thank K. Ramola for illuminating us about Lambert's function. We thank
S. Ganguly,
S. Karmakar,
G. I. Menon, 
S. Shenoy, 
K. Bhattacharya, 
M. Barma,
G. Biroli, 
and S. Sastry for many useful discussions.

\end{document}


\title{Supplementary Material: Reddy {\em et al.}}
\author{V. S. Reddy, P. Nath, J. Horbach, P. Sollich, S. Sengupta}
\maketitle

\section{The hidden first order transition: a brief background}
The background for the first order ${\mathscr N} - {\mathscr M}$ phase transition is given in Ref.~18 of the main text, henceforth cited as Nath {\em et al.} Full knowledge of this background is not necessary to appreciate the nucleation theory given in the main text -- merely assuming that such a transition exists at $\varepsilon = 0$ is enough. We refer the interested reader to existing literature \cite{sas1,sas2,sas3,sas4,sas5,sas6,sas7,popli} for further details. However, for the sake of completeness, we include below a brief summary of the background literature. 

Usually, thermal atomic fluctuations are classified as being either ``smooth" or ``singular". The former comprise long wavelength, smooth variations of the displacements, while the latter are defects, with discontinuous displacements. Our starting point is a different way of classifying atomic displacements in solids, which we hope may have more general applicability. The ${\mathscr N} - {\mathscr M}$ transition follows from this formulation. Its starting point is the realization that any given set of displacements of an atom and its neighbors within a chosen coarse graining region may be decomposed into two mutually orthogonal sub-spaces using a projection formalism~\cite{sas1,sas2,sas3}. The {\it affine} component of these displacements contains all linear transformations of a reference configuration within the coarse graining volume. These are isotropic expansion, shear strains and local rotations; trivial uniform displacements are removed by considering displacements relative to the central atom. A general displacement pattern cannot be captured completely by these linear transformations alone so that a {\it non-affine} component remains. This non-affine component may represent atomic shuffles, rearrangements, defects, slip, stacking fault~\cite{popli} etc.\ -- in other words, any displacement of atoms that is not purely elastic.  

Consider, for example, a region $\Omega$ containing a particle $i$ and its neighbours, which we label $j = 1, \ldots, n_\Omega$. Let the reference positions of the particles (say the ideal lattice sites) be given by ${\bf R}_i$. We wish to describe the instantaneous position ${\bf r}_i$ as a ``best fit'' affine transformation $\mathcal{D}$ of the neighborhood $\Omega$. This is accomplished by minimizing the error
$$\chi_i=\min_{\mathcal{D}}\left(\sum_{j=1}^{n_{\Omega}}\left(\mathbf{r}_j - \mathbf{r}_i - \mathcal{D}[\mathbf{R}_j - \mathbf{R}_i]\right)^2\right)$$
Therefore $\chi_i$ is a measure of the {\em non-affinity} at the given lattice site $i$.

By construction, the affine part $\mathcal{D}$ and the remaining non-affine part of the displacements are linearly independent~\cite{sas1}. Thermodynamically conjugate fields enhance or suppress each part independently of the other. The affine displacements couple to local stresses (and torques) while the non-affine component of the displacement averaged over all $N$ lattice sites of an ideal crystal, viz.\ $X = N^{-1}\sum_{i = 1}^N \chi_i$, couples to a new ``non-affine'' field $h_X$~\cite{sas2}. Enhancing non-affine fluctuations by increasing temperature, applying large strains or the non-affine field leads to the creation of defects~\cite{sas2,sas5}. Atomic fluctuations that act as precursors to the formation of dislocation dipoles have been shown to be the most prevalent non-affine displacement~\cite{sas2,sas5,popli}.

Nath {\em et al.} analyzed the yielding problem in the full $h_X - \varepsilon$ plane showing that the {\em equilibrium} probability distribution $P(X)$ can become bimodal for choices of $h_X$ and $\varepsilon$. While the peak of $P(X)$ at small $X$ corresponds to a rigid solid, the peak at larger $X$ corresponds to a solid in which lattice planes slip past each other to relieve stress. We called these the ${\mathscr N}$ and ${\mathscr M}$ phases respectively. It was also shown that the local stress $\sigma$ decreases in the ${\mathscr M}$ phase and coexisting ${\mathscr N}$ and ${\mathscr M}$ phases lie on either side of an interface where $\sigma$ shows a smooth step like variation. The locus of all $h_X$ and $\varepsilon$ where the areas under the two peaks are equal is the first order phase boundary. This boundary is plotted schematically  in Fig.~1 in the main text. Accurate finite size scaling showed that in the thermodynamic limit, this boundary passes through the origin so that for realistic situations ($h_X = 0$) the ${\mathscr N}$ phase is always metastable and in equilibrium, a rigid solid is non-existent. The energy of the ${\mathscr N}$-${\mathscr M}$ interface $\gamma_s$ remains well defined and finite in the thermodynamic limit all along the phase boundary.   

\subsection{Classical nucleation theory}
\begin{figure}[h]
\centering
\includegraphics[width=0.6\linewidth]{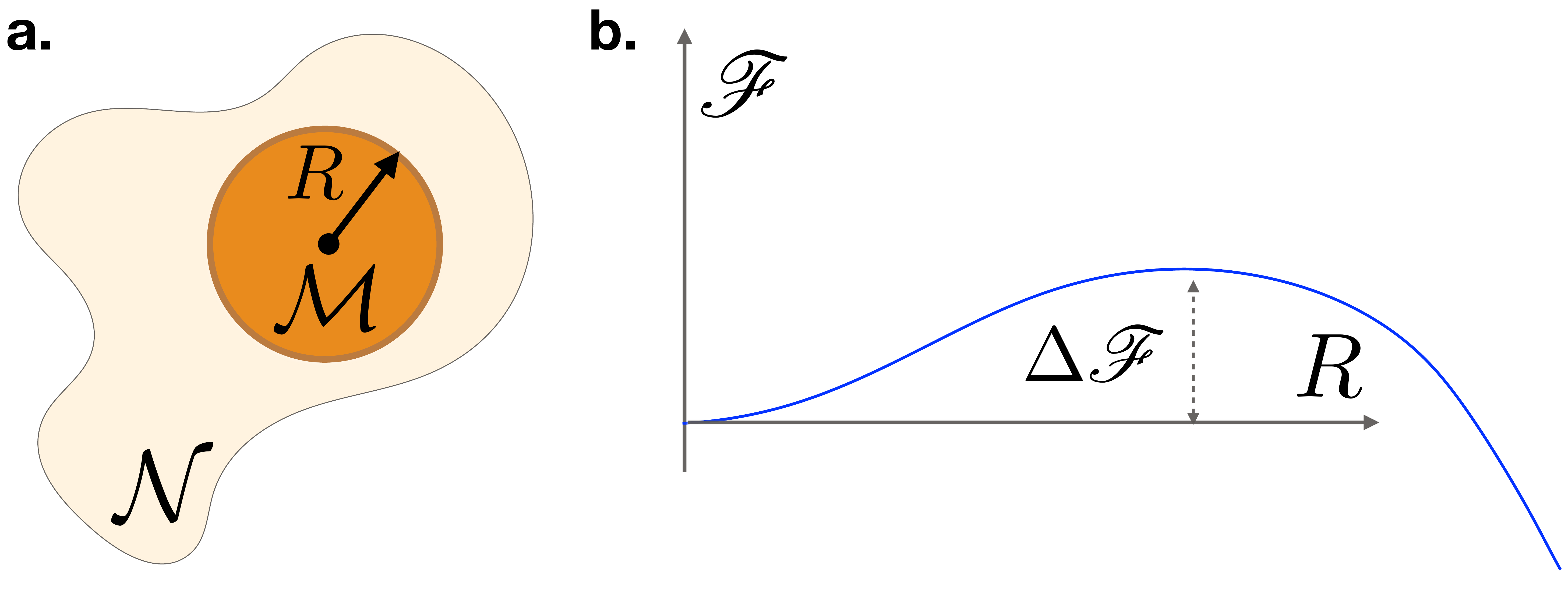}
\caption{{\bf a.} Sketch of a nucleus (bubble)  of the ${\mathscr M}$ phase of radius $R$ surrounded by the ${\mathscr N}$ phase. {\bf b.} Plot of the free energy ${\mathscr F}$ as a function of the radius of the bubble. The free energy barrier $\Delta {\mathscr F}$ is shown.}
\label{bubble}
\end{figure}

At any value of the strain, $\varepsilon \gtrsim 0$, the rigid solid becomes metastable, and bubbles of the ${\mathscr M}$ phase can form due to thermal fluctuations. In Fig.~\ref{bubble}{\bf a} we have sketched such a bubble of radius $R$. The bulk energy of formation of this bubble is always negative since the ${\mathscr M}$ phase cannot support stress and has zero elastic energy. However, in order to form the bubble, an  interface with positive free energy $\gamma_s$ needs to be formed. The total free energy of the bubble increases for small $R$ but decreases when $R > R_c$, where $R_c$ is the critical radius (see Fig.~\ref{bubble}{\bf b}). The nucleation barrier $\Delta {\mathscr F} = {\mathscr F}(R_c) - {\mathscr F}(0)$. Spontaneously formed bubbles with $R < R_c$ shrink and those with $R > R_c$ grow to form the new phase. The mean first passage time is the average time that the system has to wait before a spontaneous fluctuation with $R > R_c$ appears. This, at any temperature $T$, is given by the Kramer's formula $\tau_{FP} \propto \exp(\beta \Delta {\mathscr F})$.
In the main text we have shown that the nucleation barrier $\Delta {\mathscr F} \sim \varepsilon^{-4}$ in three dimensions. This means that as a solid is deformed at a continuous rate, the barrier also keeps decreasing. In this case the Kramer's formula gets modified into a self consistency problem as we show below. 

The self consistent nucleation theory used in our work has been derived by Shillcock and Seifert in Ref. 30 of the main text (henceforth cited as SS) in the context of the escape of a particle from a metastable minimum in one dimension, under the influence of a time-ramped force $f(t) = \mu t$. Following closely the notation of SS we write the Langevin equation for the position of the particle $x(t)$, 
$$ \partial_t x = - \partial_x V(x) + f(t) + \zeta,$$ where the noise $\zeta$ has the usual Gaussian statistics with $\zeta(t)\zeta(t') = 2 \delta(t-t')$. The potential $V(x)$ has a minimum at $x_m$ and a saddle point at $x_s$ and the barrier $Q(f) = Q_0 - f(x_s(f) - x_m(f))$ and $Q_0 \equiv V(x_s) - V(x_m)$. One can show that the mean first passage time $T_0(x,f)$ for a particle at $x$ escaping the metastable minimum at $x_m$ in the presence of a constant force $f(t) = f$ is given by the solution of the following differential equation, $$ (f\partial_x + {\mathscr L}^{\dagger})
T_0(x,f) = -1,$$ with the backward Fokker-Planck operator ${\mathscr L}^{\dagger} \equiv (-\partial_x V \partial_x + \partial^2_x)$. This differential equation can be solved formally and for large $Q(f)$; the explicit (Kramer's) solution may be written down as$$ T_0^K(f) = \tau_0(f) \exp[Q(f)],$$ where the characteristic time or inverse ``attempt frequency",  $\tau_0(f) \equiv 2\pi/\sqrt{|\partial^2_xV(x_s)|\partial^2_xV(x_m)}$. The self consistency approximation for a time varying force $f(t)$ amounts to assuming that $f$ is dominated by its value at $T_0(x,f)$, which leads to a self consistent relation for $\tau_{FP} = T_0(x,\mu \tau_{FP})$.  For time independent barriers, $\tau_0$ is a constant but, in general, depends on $f$ through the $f$ dependence of $x_m$ and $x_s$.    

We have adapted this solution for the classical nucleation theory problem with time dependent barriers -- a much more complex problem. In the simple one dimensional model of SS shown above the attempt frequency, i.e.~the inverse of the characteristic time $\tau_0$, is given in terms of the curvatures of the potential energy surface at the minimum and the saddle point. In this case, the attempt frequency depends on the time dependent force that changes the barrier.

In the nucleation problem, the characteristic time $\tau_0$ is connected to microscopic processes that produces the bubble fluctuations. For example, in the case of nucleation of liquid droplets from gas, it is a microscopic atomic diffusion time. In our case it should be related to the nucleation of dislocation pairs or loops. As reported in Nath {\em et al.}, $\tau_0$ becomes independent of the barrier height for small ${\dot \varepsilon}$ when $\Delta {\mathscr F}$ is large. For faster rates, $\tau_0$ increases as the lattice itself becomes less stable (flatter potential energy surface), barriers tend to vanish and the instability regime is approached. Our theory, as well as the Kramer's approximation itself, becomes invalid in that limit. 

\subsection{Influence of pre-existing defects}
Our theory, as explained in the main text, describes yielding for the situation where the rigid solid is metastable, with reasonably large $\beta \Delta {\mathscr F} \gtrsim 1$. We have mentioned that this limit corresponds to having only a small number of pre-existing dislocations. How does our theory break down as the ideal solid is gradually disordered by incorporating more and more defects?    
\begin{figure}[h]
\centering
\includegraphics[width=0.7\linewidth]{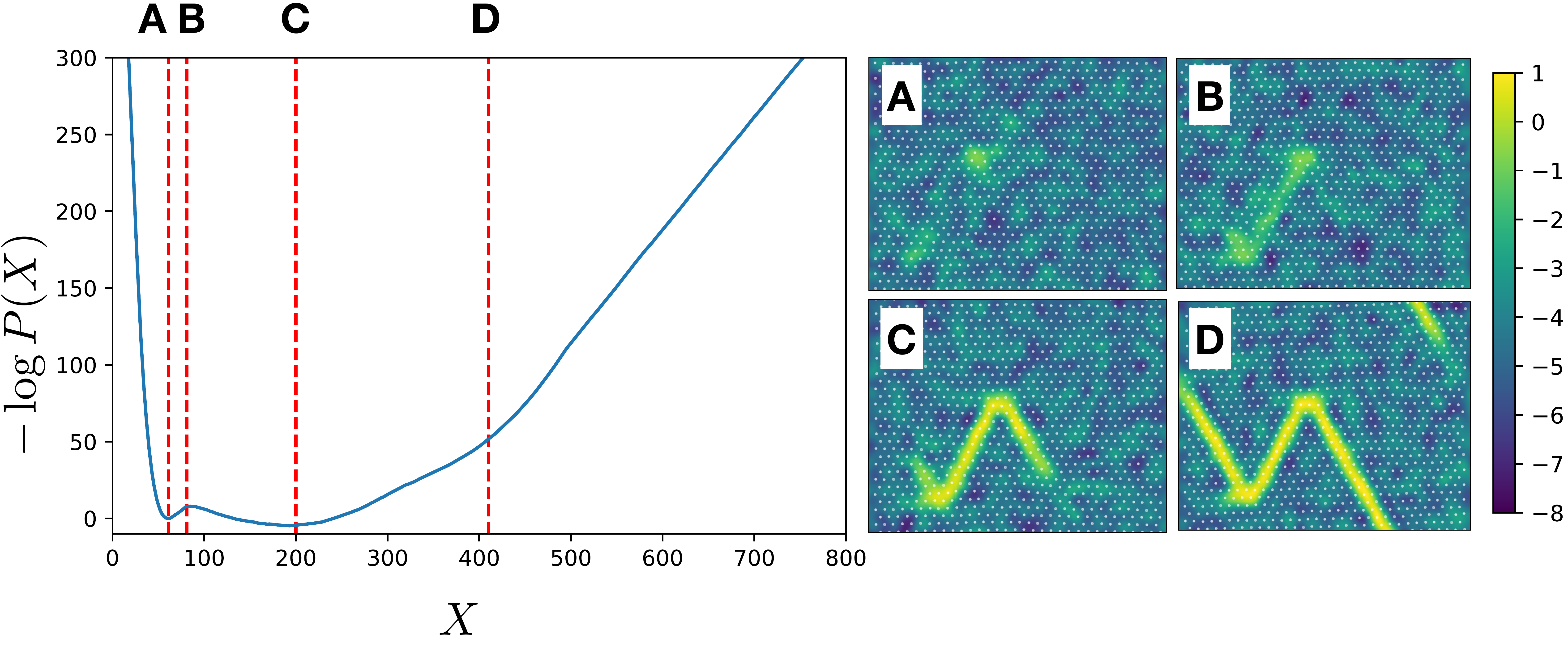}
\caption{The probability $P(X)$ at coexistence ($h_X=-5$, $\varepsilon=0.02$), for $N=1014$ Lennard Jones particles at reduced temperature $T = 0.8$ and density $\rho = 2/\sqrt{3}$. Configurations for four values of $X$ (red dashed vertical lines) $A-D$ are also shown.}
\label{defect}
\end{figure}

This is a hard question to answer although an estimate can be made from numerical calculations. For this, we repeated the equilibrium calculation of $P(X)$ in Nath {\em et al.} with a reference structure that now contains a single dislocation pair in a system of $N \approx 1014$ Lennard Jones particles arranged otherwise as a triangular lattice in two dimensions. The dislocation is introduced by deleting $10$ particles in a close packed lattice row in a $32\times32$ lattice and minimizing the energy. Our result for $P(X)$ close to phase coexistence is  shown in Fig.~\ref{defect}. This defect density $\sim 0.2\%$ is rather large. Smaller defect concentrations can only be achieved by going to larger systems. Considering that the sequential umbrella sampling Monte Carlo technique used here (see Nath {\em et al.} for details) is quite complex, systems much larger than this become prohibitive to handle. 

Comparing with the ideal case reported in Nath {\em et al.}, we observe that the relative height of the saddle between the two minima in $-\log P(X)$ is about $20$ times lower than in a comparable ideal solid. Configurations show that the new phase now nucleates at the location of the existing defect i.e. heterogeneous, instead of the homogeneous nucleation seen in Nath {\em et al.}. This results in a lower $\gamma_s$, the interfacial energy and ultimately the nucleation barrier $\Delta {\mathscr F}$, which should also be correspondingly lower. Further, additional metastable configurations corresponding to extra local minima in $-\log P(X)$ appear and the equilibrium phase boundary shifts to lower $\varepsilon$ values implying that the rigid ${\mathscr N}$ solid is less thermodynamically stable. It is conceivable that as the defect concentration is increased even further, the nucleation barrier will  eventually disappear. It is in this limit that our theory will cease to be valid.

\section{Brief summary of data sources and fitting procedures}
In this section, we briefly summarize, for the sake of completeness, the experimental and simulation results which we use in the main text. This is followed by a note on our fitting procedure. We discuss below each of our sources in the order that they appear in Table 1 and using the reference numbers from the main text. We are of course limited by the information shared by the original authors. For example, in some of the experiments and simulations the temperature is not mentioned (we have assumed room temperature i.e. 300 K). For other details, such as the exact experimental protocols or expressions for the interatomic potentials we refer the reader to the original papers. 

In Ref.~32, the authors have performed uniaxial loading experiments on single-crystalline Cu nano-pillars of diameters ranging from 75 to 500 nm to measure its strain-rate sensitivity. The Cu nano-pillars are fabricated by electroplating Cu from $\rm{CuSO_4}$ solution. The loading axis of the nano-pillars is orientated along the $\rm{\langle 111 \rangle}$ direction. The authors have also reported the variation of surface roughness to be $\rm \pm 2\,nm$ which resulted in an  error of $\rm \pm 4\%$ in the measurement yield strength for 100 nm pillars. Their TEM analysis estimated a dislocation density of $\rm \sim 10^{14} m^{-2}$ in the prepared samples.

In Ref.~33, the authors have studied the tensile properties of nano-crystalline Cu prepared from 99.999\% pure Cu powders via a two-step ball milling process, initially at cryogenic temperatures ($\rm LN_2 T$), followed by milling at room temperature. Their TEM images show that the grains are mostly equi-axed with an average size of 54 nm. With these samples, the authors have performed tensile tests at two different (77$^\circ$ K and 300$^\circ$K) temperatures and studied the strain-rate sensitivity and major deformation modes in detail.

In Ref.~34, the authors have performed \textit{in situ} uniaxial tensile experiments with a micro-mechanical device in order to quantitatively measure the strain rate dependence of 0.2\% offset yield stress in single-crystal nickel nano-wires with diameters ranging from 80 to 300 nm and strain-rate varying from $\rm 10^{-4}$ sec$^{-1}$ to $\rm 10^{-2}$ sec$^{-1}$. The single-crystal Ni nano-wires were prepared via an electrochemical deposition method with nano-porous anodic aluminum oxide templates with pore sizes ranging from 100 to 300 nm. Their TEM analysis shows that the nano-wires with diameter of $\sim$100 and $\sim$200-300 nm are single-crystalline with $\rm \left[111 \right]$ and $\rm \left[112 \right]$ axial orientation, respectively.

In Ref.~35, the authors have performed systematic experiments to measure the SRS of deformation in fully dense nano-crystalline Ni using two independent experimental techniques, viz. depth-sensing indentation test and tensile tests. The Ni samples were prepared using electrodeposition with a thickness of 150 $\mu$m. The authors investigated Ni nano-crystals of a range of grain sizes. While nano-crystalline (40 nm) Ni showed SRS, this effect is absent in Ni samples with larger grain sizes.

In Ref.~36, in close resemblance with Ref.~26, the authors have investigated SRS of nano-crystalline Ni samples of 20 nm mean grain size using nano-indentation and tensile testing at room temperature. They report that nano-crystalline Ni exhibits higher strain-rate sensitivity during nano-indentation than during tensile testing. The Ni samples were prepared from a sheet of fully dense nano-crystalline Ni of thickness $\rm \sim 400\, \mu m$ using a surfactant-assisted direct-current electrodeposition technique. Their XRD results show that the deposited Ni has a preferential orientation along the $\left[200 \right]$ planes.

In Ref.~37, the authors have performed similar experiments to investigate SRS of crystalline Ni samples of different sizes ranging from macro to micro to nanometer scale for over eight orders of magnitude change in strain-rate. The authors use tensile testing and micro-scratch testing to evaluate SRS of the tensile strength and the scratch hardness variation with scratch velocity, respectively. They report that the SRS increases with decreasing grain size.

In Ref.~38, the authors have performed MD simulation of Gold nano-wires under tensile loading at  
room temperature with strain rates varying from $\sim 10^{7}$ to $10^{10}$ sec$^{-1}.$ The simulations were carried out by taking a square-cross-section gold [100] nano-wires of length 16 nm out of a bulk fcc crystal. The system was relaxed to an equilibrium minimum energy configuration with free boundary conditions, followed by a thermal equilibration to 300 K for 20 ps using a Nos\'e-Hoover thermostat. The authors used two different forms of embedded atom model (EAM) potential (Voter-Chen and Foiles) to model the interatomic interactions and arrived at similar results.

As in Ref.~38, the authors of Ref.~39 study the similar effect of tensile loading of gold nanowire at different strain rates. The initial structure was created by choosing a cylindrical cutoff radii from a fcc single crystal of Au which resulted in a nanowire having initial diameter and length of 5.65 nm and 22.6 nm, respectively. Then the system was thermally relaxed to 273$^\circ$K for 50 ps using Nos\'e-Hoover thermostat. A Berendsen barostat was also employed to control the pressure in each direction. The interatomic interactions were modeled by an EAM potential. Then the study investigated deformation mechanism with aid of stress-strain curves at constant strain rates ranging from $\sim 10^{8}$ to $10^{10}$ sec$^{-1}$. 

In Ref. 40, the authors have studied the effect of strain-rate dependent tensile properties of the helical multi-shell (HMS) gold nano-wires using MD simulations. The system was prepared by rolling up a Au (111) crystal with a certain helical angle producing a cylindrical wire consisting of concentric shells of Au atoms. The many-body tight-binding potential was used to model the interatomic interactions and the system was kept at constant temperature of 4K by re-scaling the velocities of the particles. The study reports deformation properties and yielding behavior of HMS gold nano-wires of different structures, viz. 7-1, 11-4, 14-7-1; the numbers referring to the number of Au atoms in each shell of the circular cross section starting from the outermost, for a range of strain-rates ranging from $\sim 10^{7}$ to $10^{11}$ sec$^{-1}$. We have used the data for the thickest (14-7-1) wire.

In Ref.~41, the authors have performed a MD study of the mechanical properties of Ni nano-wires for a strain-rate range of $\rm \sim 10^8 \,to\, 10^{11}\,$ sec$^{-1}$ at room temperature. The study uses quantum corrected Sutten-Chen (Q-SC) type many-body force field to model the interatomic potential. The initial structure having a diameter of 2.53 nm was constructed by  cutting off a cylinder centered at a cubic interstitial site of a large cubic fcc single crystal of Ni. The system was thermally relaxed to 300 K using Nos\'e-Hoover thermostat for $5\times 10^4$ time steps. Then the authors performed tensile loading experiments at different strain rates and report the mechanical behavior.

In Ref.~42 and 43, the authors have simulated the effect of uniaxial tensile strain on a single crystal Cu nanowire in order to find its strength and mechanical properties. The initial structure was prepared from a bulk single crystal copper and was equilibrated to a constant temperature using direct velocity rescaling method. The EAM potential model was used. Then the authors report yield stress, elastic constants and other mechanical properties for a range of strain-rates spanning from $\rm \sim 10^{7}\,to\,10^{11}$ sec$^{-1}$.

SRS is a weak effect; changing the deformation rate by orders of magnitude makes only a small difference in the yield stress. Therefore both the experiments and simulations are demanding and care needs to be taken in order to obtain a signal over and above random noise. In spite of that, the data continues to have a fair amount of scatter, with some data sets affected more than others. As we have mentioned in the main text, all of this data was fitted to power laws in the original papers. The data itself is therefore not accurate enough and is not available in a wide enough range to be able to distinguish between alternate theories. The point of our work is to show that our theory, which is consistent with the rigorous thermodynamic requirement of shape independence of equilibrium free energies is {\em also} able to describe the data equally well. 
\begin{figure}[t]
\centering
\includegraphics[width=0.5\linewidth]{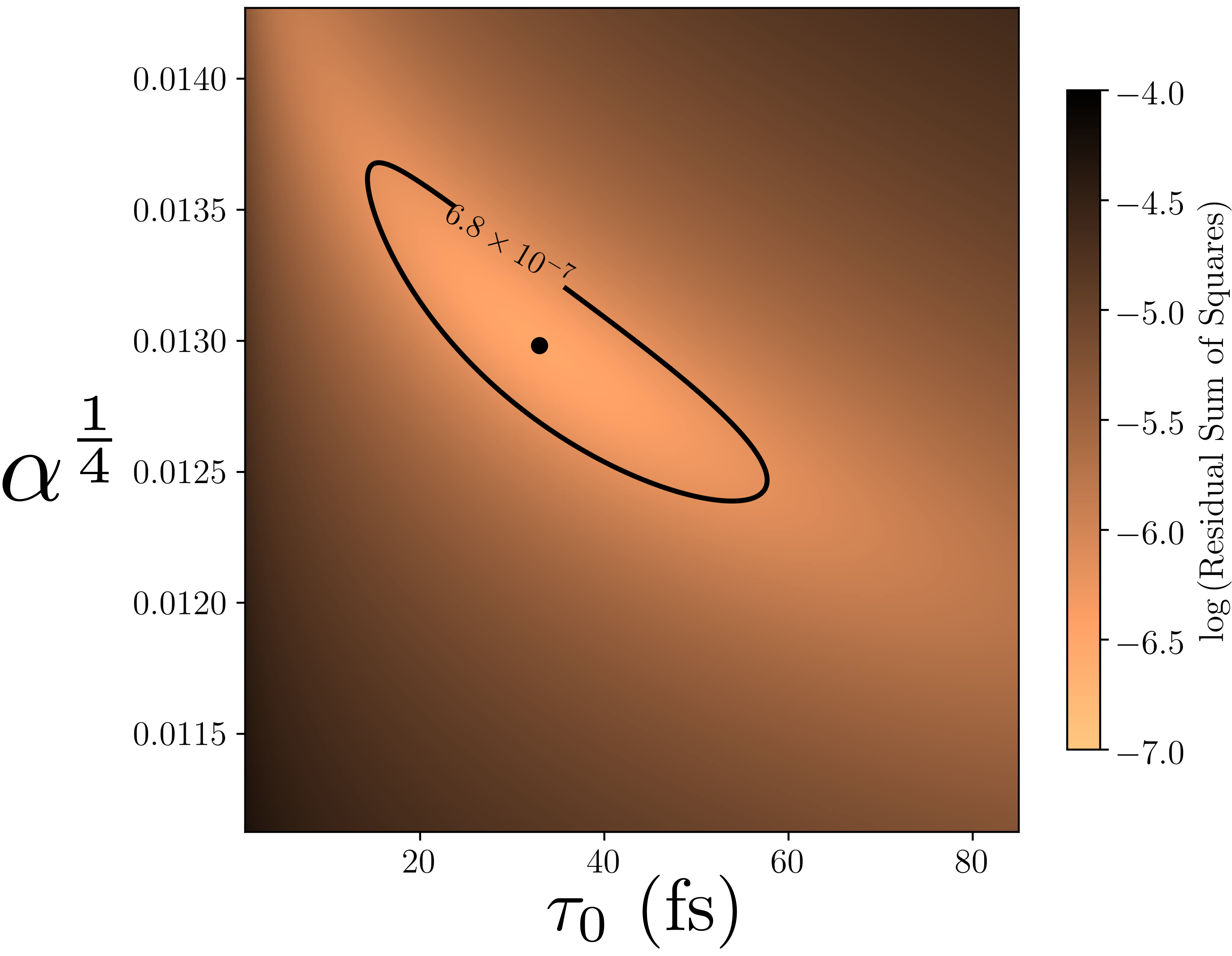}
\caption{Contour plot of the residual sum of least square errors for the fit to the two parameters $\alpha^{\frac{1}{4}}$ and $\tau_0$. The contour for the $95\%$ confidence level is shown corresponding to an error of $6.8\times10^{-7}$. The black dot in the centre corresponds to minimum error and is the value quoted in Table 1 of the main text.}
\label{fit_confidence}
\end{figure}

For most of the fits shown in Table 1 of the main text, we use the relatively crude, but standard, Nelder-Mead, downhill simplex, method. For Fig.~2f we needed more accurate fits of data from experiments and simulations on the same material and similar protocol, ensuring at the same time that both data sets have relatively small scatter. Unfortunately, only a single pair, namely, Ref.~33 from experiments and Ref.~43 from MD on Cu, satisfied our requirements. For this particular fit, we use another standard algorithm, viz.~Levenberg-Marquardt, for fitting the simulation data, which here is smooth and available over a relatively large range of strain rates. The fitted values for $\alpha$ and $\tau_0$ turned out to be similar (within the error of both the fits) to the earlier downhill simplex fits. We repeated this for other data sets as well (not shown), in each case obtaining similar fit parameters depending on the quality of the data. For the fit used in Fig.~2f, we plot in Fig.~\ref{fit_confidence} the residual sum of squares as a function of the two fitting parameters obtained by fitting the data for yield stress as a function of strain rate from Ref.~43. The uncertainty in the fitted value of $\alpha^{\frac{1}{4}}$ is  $\lesssim 10\%$, which is the same for the interfacial energy $\gamma_s$ while $\tau_0$ has a larger, $\sim 20\%$ error. 

Is our theory falsifiable or does it have enough parameters to fit any possible set of data? We have only two parameters in our theory. These parameters merely scale the fitting function Eq.~(1) (main text) along the two axes ($\varepsilon^*$ and $\dot \varepsilon$) as is clear from Eq. (2) (main text). The shape of the $\varepsilon^*(\dot \varepsilon)$ curve, a monotonic function passing through the origin, does not change under this scaling. There is no separate intrinsic scale for the yield point - no ``rate independent reference mechanical threshold" - and yet Eq.~(1) (main text) is able to fit such a large body of data purely due to the correct functional form of the nucleation barrier and the nature of its singularity at zero deformation that follows from our derivation. We find this quite remarkable. To make this clearer, we show in the next section a comparison of our theory to the qualitatively similar work presented in Ref.~17 of the main text. 


\section{Comparison with Sausset, Biroli and Kurchan}
Our work, both the present one and that reported in Nath {\em et al.}, has been, to a large extent, inspired by Sausset, Biroli and Kurchan (Ref. 17 of main text, henceforth cited as SBK). SBK showed that a crystalline solid is able to flow when subjected to any external stress, however small. The viscosity of the crystal, unlike a liquid, is singular and diverges as stress approaches zero. Similar to the calculation presented in the main text, SBK considered nucleation of stress-free regions within a rigid solid and obtained the critical size of these droplets, the barrier to nucleation and the mean first passage time $\tau_{FP}$. While the two calculations, SBK on the one hand and that of Nath {\em et al.} and the present work on the other, are similar in spirit, they differ in one very important aspect. Unlike the present calculation, SBK did not consider the flow process as the decay of a stressed metastable solid to another distinct stable phase following a quench across a first order phase transition. The barrier is calculated considering the nucleation of dislocation loops wholly contained within the rigid solid. The main consequence of this is that, in SBK, the calculated interfacial energy of the droplet, $\gamma_s$, vanishes as the stress $\sigma$ tends to zero. This, in turn, leads to a free energy barrier $\Delta {\mathscr F}$ that has a weaker divergence as $\sigma \to 0$ than the one derived in our work, where $\gamma_s$ is a non-vanishing constant. 

In SBK, no quantitative assessment of the results by comparison with simulations or experiments was given. We do this below beginning 
with Eq.~(4) of SBK:
\begin{equation}
\tau_{FP} = \tau_0 \exp\left[ \frac{k \sigma^c_y}{\sigma} \left(-\frac{1}{8} \ln\frac{\sigma}{\sigma^c_y} \right)^3 \right],\label{eq1}
\end{equation}
where the quantity $\sigma^c_y = K^2 a^3/k_BT$, $K$ is an elastic modulus, $k$ is a number of $\sim \mathcal{O}(1)$, $a$ the lattice parameter and $k_B T$ the thermal energy. First, we write the above relation in terms of the strain $\varepsilon = K^{-1} \sigma$ with $\varepsilon_y^c = K a^3/k_BT$, so that Eq.~\eqref{eq1} now reads
$$\tau_{FP} = \tau_0 \exp\left[ \frac{k \varepsilon^c_y}{\varepsilon} \left(-\frac{1}{8} \ln\frac{\varepsilon}{\varepsilon^c_y} \right)^3 \right].$$
Multiplying both sides by $\dot{\varepsilon}/{\varepsilon_y^c}$ gives
$$\frac{\dot{\varepsilon}\,\tau_{FP}}{\varepsilon_y^c} = \frac{\dot{\varepsilon}\tau_0}{\varepsilon_y^c} \exp\left[ \frac{k \varepsilon^c_y}{\varepsilon} \left(-\frac{1}{8} \ln\frac{\varepsilon}{\varepsilon^c_y} \right)^3 \right].$$
Finally, identifying the yield point $\varepsilon^* = \dot{\varepsilon}\,\tau_{FP}$, we obtain a self-consistency relation in $\ell =\ln(\varepsilon_y^c/\varepsilon^*)$;
%
 \begin{align}
     -\ln  \dot{\tilde{\varepsilon}} = \ell + \frac{k}{512}\; \ell^3\,e^{\ell}, \label{eq2solve}
 \end{align}     
where, $\dot{\tilde{\varepsilon}} = \dot{\varepsilon}\tau_0/\varepsilon_y^c$ is the rescaled deformation rate.
The solution can then be converted to the function $\varepsilon^* = \varepsilon^c_y \, e^{-\ell}$ vs.\ $\dot \varepsilon =  \dot{\tilde{\varepsilon}}\, \varepsilon_y^c/\tau_0$ to compare, if needed, with the experimental and MD results. We now give below a set of realistic parameters for $\tau_0, \dot{\varepsilon}, K$ etc. We use the values for Cu (1), i.e.~MD simulations of a $9.18\times9.18$ nm nano-wire from the main text,
\begin{center}
{\renewcommand{\arraystretch}{1.2} 
\begin{tabular}{ |c|c| } 
 \hline 
 Quantity & Value \\
 \hline\hline
 $\tau_0$ & $36.59 \; \mathrm{fs}$  \\ 
  \hline
 $\dot{\varepsilon}$ & $10^{-4} \; {\rm to}~ 10^{11}  \;\mathrm{sec^{-1}}$  \\ 
  \hline
 $K$ & $130 \; {\rm GPa}$  \\ 
  \hline
 $a$   & $3.597 \; \angstrom$ \\
  \hline
 $k_BT$ & $4.11 \times 10^{-21} \; \mathrm{J}$ \\
 \hline
\end{tabular}
}
\end{center}
We have therefore
\begin{align}
\varepsilon^c_y = \frac{K a^3}{k_B T} &= \frac{130 \times 10^9 ({\rm J/m}^3) \times (3.597 \times 10^{-10} {\rm m})^3}{4.11 \times 10^{-21} {\rm J}} \nonumber \\
&=1.472 \times 10^3 \nonumber
\end{align}
The possible range of $ \dot{\tilde{\varepsilon}}$ comes out to be~:
\begin{align}
 {\dot{\tilde \varepsilon}}  &= (10^{-4} \ldots 10^{11})\times \frac{3.659 \times 10^{-14}}{1.472\times 10^3} \nonumber \\
   &= 2.48\times 10^{-21}  \ldots 2.48\times 10^{-6} \nonumber 
   \sim 10^{-21} \ldots 10^{-6}. \nonumber
\end{align}
This implies  a range of $\sim -48 \ldots -13$ for the logarithm,  $ \ln \dot{\tilde{\varepsilon}}$.
%

\begin{figure}[h]
\centering
\includegraphics[width=0.7\linewidth]{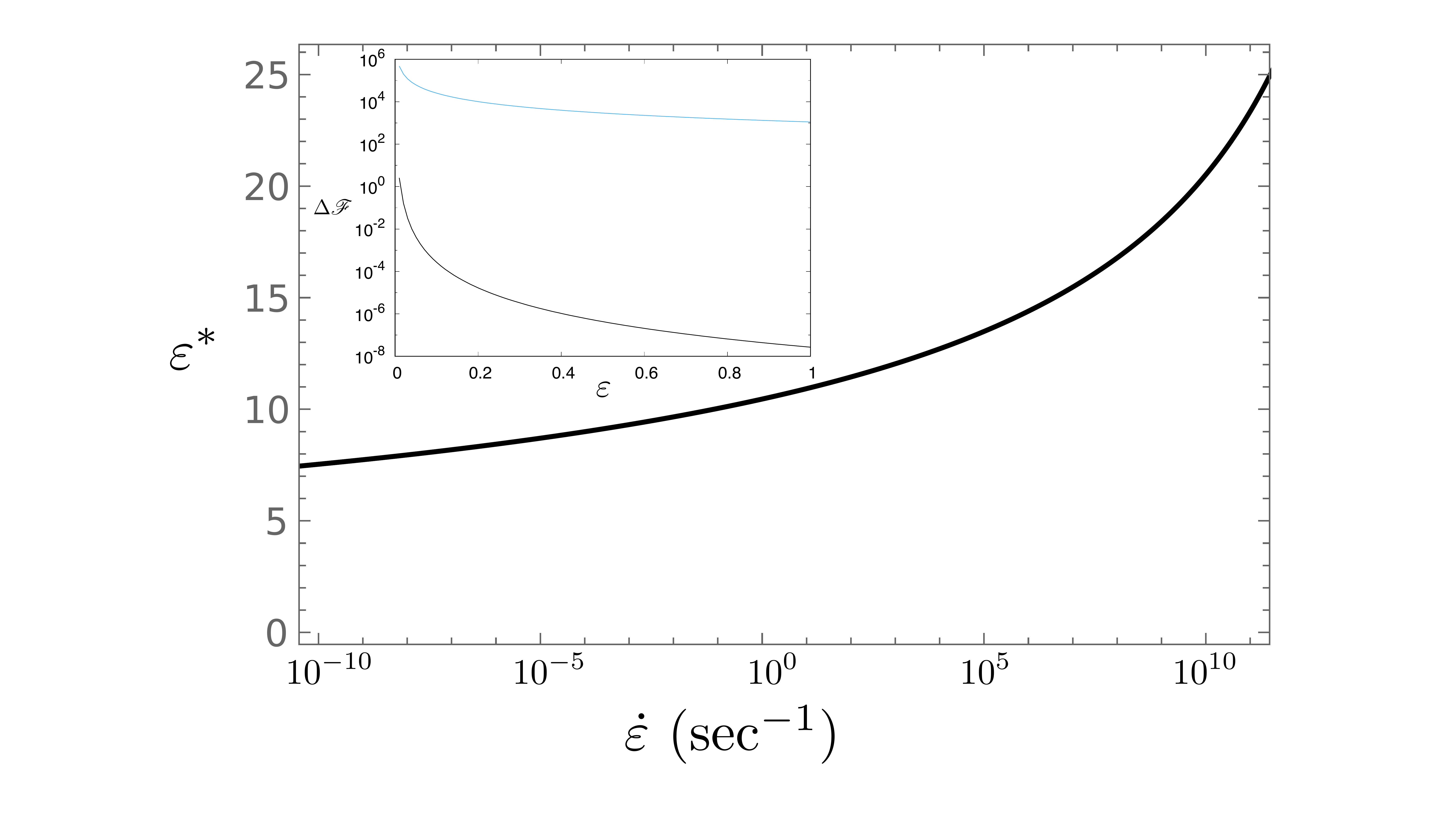}
\caption{$\varepsilon^*$ as a function of $\dot{\varepsilon}$ with $k=1$ and $\tau_0=36.59\; \mathrm{fs}=3.659\times 10^{-14}\;\mathrm{sec}$ obtained by numerically solving Eq.~\eqref{eq2solve}. The inset shows the nucleation barrier $\Delta {\mathscr F}$ obtained from our work (black line) and from that of SBK (blue line).}
\label{k1tauCu}
\end{figure}

Unfortunately, Eq.~\eqref{eq2solve} cannot be solved analytically but numerical solutions are readily obtained. Solving for $\ell$ over the range of values of $\dot{\tilde{\varepsilon}}$ we obtain  $\varepsilon^*$ as a function of $\dot{\varepsilon}$ in sec$^{-1}$ in Fig.~\ref{k1tauCu}. We have used $k=1$ and $\tau_0=36.59 \;\mathrm{fs}\, (=3.659\times 10^{-14}\;\mathrm{sec})$ for our calculation.
It can be clearly seen that, although the curve has the same shape as that shown in Fig.~2f of the main text, the yield strain obtained with $k=1$ is substantially larger than is plausible physically. Indeed, to obtain meaningful values of $\varepsilon^*$ one either needs to go much below the deformation rate corresponding to the slowest available experiments or change the values of $k$ and $\tau_0$ drastically and beyond physically reasonable limits; small changes in the parameters do not change the results shown. This is also clear from the plot of the nucleation barrier $\Delta {\mathscr F}$ (Fig.~\ref{k1tauCu}(inset)). The SBK description obtains a barrier that has a weaker singularity as $\varepsilon \to 0$, but the numerical value is larger for physically relevant $k$. 
%

This calculation actually illustrates a very important aspect of the theory presented in both our work and that of SBK. The two approaches are similar in that there are no adjustable, rate-independent mechanical thresholds at finite temperatures and unlike some phenomenological models, they are both falsifiable. The $\varepsilon^*({\dot \varepsilon})$ curve depends solely on the nature of the singularity of $\Delta{\mathscr F}$ in the $\sigma \to 0$ limit. The asymptotic form derived in SBK, however, fails to quantitatively explain the strain rate dependence of the yield point. It is entirely possible that a more accurate expression may give a better result, although at present there is no quantitative evidence for this. In contrast our theory, based on the hidden first order transition scenario described in Nath {\em et al.}, is able to describe the data using physically reasonable values of the parameters viz.~$\tau_0$ and $\gamma_s$. 
\bibliographystyle{unsrt}